\documentclass[]{spie}  %>>> use for US letter paper
\usepackage{amsmath,amsfonts,amssymb}
\usepackage{graphicx}
\usepackage{graphics}
\usepackage{subfigure}
\usepackage{caption}
\usepackage{subcaption}
\usepackage[colorlinks=true, allcolors=blue]{hyperref}
\usepackage[dvipsnames]{xcolor}
\usepackage{wrapfig}
\usepackage{url}

\title{SPIE Proceedings: GALI - A GAmma-ray burst Localizing Instrument: Results from Full Size Engineering Model}
\author[a]{Julia Saleh-Natur}
\author[a]{Ehud Behar} 
\author[a]{Omer Reich}
\author[a]{Shlomit Tarem}
\author[a]{Zvika Tarem}
\author[a]{Alex Vdovin}
\author[a]{Amir Feigenboim}
\author[b]{Roi Rahin}
\author[c]{Avner Kaidar}
\author[c]{Hovhannes Agalarian}
\author[d]{Alon Osovizky}
\author[d]{Max Ghelman}
\affil[a]{Department of Physics, Technion - Israel Institute of Technology, Haifa, Israel}
\affil[b]{NASA Goddard Space Flight Center, Greenbelt, Maryland, NASA}
\affil[c]{Asher Space Research Institute, Technion, Haifa, Israel}
\affil[d]{Nuclear Research Center, Negev,Israel}
\authorinfo{Further author information: (Send correspondence to J.S.N.)\\J.S.N.: E-mail: Julia.s@campus.technion.ac.il }%, \\  R.R.: E-mail: roi.a.rahin@nasa.gov}

\begin{document} 
\maketitle

\begin{abstract}

We present a full-size engineering model of GALI - The  GAmma-ray burst Localizing Instrument, composed of 362 CsI(Tl) small cubic scintillators, distributed within a small volume of $\sim2$l, and read out by silicon photo-multipliers. GALI can provide directional information about GRBs with high angular accuracy from angle-dependent mutual obstruction between its scintillators.
 Here, we demonstrate GALI's laboratory experiments with an $^{241}$Am source, which achieved directional reconstruction of $<$3$^\circ$ accuracy, in agreement with our Monte-Carlo simulations. GALI has a wide field view of the unobstructed sky.
 With its current cubic configuration, GALI's effective area varies between 97\,cm$^2$ (face on) and 138 cm$^2$ (from the corners at 45$^\circ$), which is verified in the current experiment. %by the highest count rate.

%250 word.
%one paragraph
\end{abstract}


\section{INTRODUCTION}
\label{sec:intro}  % \label{} allows reference to this section

\subsection{Gamma-Ray Bursts Phenomenology and Current Detecting Methods}
\paragraph{} $\gamma$-ray bursts - GRBs -  are the most luminous transient events in high-energy astrophysics with a duration of $0.1$ $\sim$ 100 seconds \cite{P.Mésźros}. Their nonthermal energy spectrum is in the soft gamma regime between several keV to several MeV \cite{Kumar2015}.  The energy spectrum of GRBs decays as a power law in energy with a photon index that varies between 1-3. These electromagnetic explosions are observed isotropically across the sky \cite{Kouveliotou1933}.
%The GRBs radiated energy is between $E_{iso}=10^{48}$ and $10^{55}$\,erg if the emission is isotropic. This is much more than the energy that core-collapse supernovae or compact star mergers radiate.  

\paragraph{} GRB sources are likely jetted. With time increasing, the energy spectrum of a GRB becomes softer, due to the cooling of the relativistic jet. Long GRBs (duration $>2$\,s) are associated with stellar core-collapse supernovae with  $M\gtrsim$ 15\,$M_{\odot}$. This is due to the spectroscopic detection of type Ic supernovae at GRB sites, mostly in star-forming regions \cite{Fruchter2006}.  Short GRBs (duration $<2$\,s) are associated with compact stellar mergers of two neutron stars or a neutron star with a black hole \cite{JANKA1999}. Several short GRBs were found in elliptical galaxies where there is an older stellar population. Short GRBs seem to be less energetic than long ones and have harder spectra \cite{Barthemly2005, Kouveliotou1933}. The wide assortment of GRB light curves emphasizes that there are no specific features or patterns that all GRBs follow \cite{Arnon2004}. Yet, there are a few common ones. For high $\gamma$ energies, the duration of the GRB peak decreases, and the peak intensity shifts to a shorter time. 

\paragraph{} Since 1973, when sixteen GRBs were first observed by detectors on the Vela satellites, many instruments and numerous publications were devoted to understanding the GRB properties. 
The main challenge today for GRB detectors is to pinpoint their direction in the sky accurately enough within a few seconds to allow for multi-wavelength follow-up. %Optimally, this capability is available for a wide field of view.
 The Swift-BAT (Burst Alert Telescope \cite{Barthemly2005}) instrument has revolutionized this field. 
It employs a coded-mask aperture that detects and localizes GRBs with an angular accuracy of 1-4 arc minutes in a field of view of 1.4\,sr. It operates in the energy range of 15-150 keV with $\sim$7\,keV resolution. The Swift-BAT detector plane is composed of 32,768 pixels of CdZnTe (CZT). One meter above it, the coded mask is composed of $\sim52,000$ pieces of lead. Fermi-GBM (Fermi Gamma Ray Observatory\cite{Meegan2009}) has 12 NaI(Tl) - thallium-activated sodium iodide - and 2 BGO - bismuth germanate - scintillation detectors. %The 12 NaI(Tl) detectors measure energies up to 1\,MeV, and are used for GRB directional localization. 
The directions of the NaI(Tl) detectors are arranged in such a manner that the directional localization of GRBs can be reconstructed from the relative count rates on each detector.
The BGO detectors detect high energy photons up to $\sim40$\,MeV. They are placed on opposite sides of the spacecraft to guarantee at least one of them will observe any burst above the horizon. GBM has an angular accuracy of 3-5 degrees, and a large field of view (full un-occulted sky of 9.5\,sr).

\subsection{GALI Concept} 
\paragraph{} 
%GRBs are extremely difficult to pinpoint in space since they last only seconds. Determining their angular location requires scanning the entire sky. Thus a wide field of view and high angular accuracy are needed to localize their sources. 
The need for high angular accuracy of GRBs combined with full sky coverage remains vital and was clearly illustrated by the first LIGO-VIRGO detection of a neutron star merger in 2017, GW170817 \cite{Abbott2017a}. 
Despite a time coincident short GRB (170817A) independently detected by the Fermi/GBM and by INTEGRAL/IBIS \cite{Abbott2017a}, it took 11\,hrs to identify and locate the host galaxy NGC\,4993 \cite{Arcavi2017}.

\begin{figure} [h]
    \centering
    \includegraphics[width=0.9\linewidth]{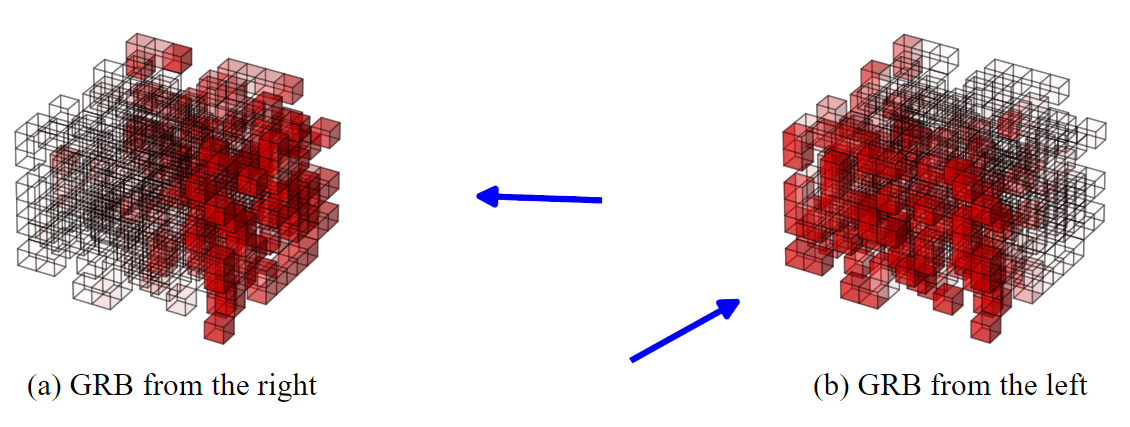}
    \caption{\label{fig:simulatuins} GALI localization concept based on two extracts from our simulations (see \autoref{sec:simulations}). The blue arrows represent the direction of a GRB. The source counts recorded in each scintillator cube are marked by shades of red, where more opaque colors represent higher count rates. The different patterns between the two directions are discerned and provide high localization accuracy.}
\end{figure}

\paragraph{} As part of the effort to increase the detecting and localizing efficiency of the current instruments, the present work introduces GALI - A GAmma-ray burst Localizing Instrument built according to a novel configuration of scintillators, which enables accurate localization of GRBs. Its working principle is based on angle-dependent mutual obstruction between the many small scintillators, which are distributed within a small volume. \autoref{fig:simulatuins} illustrates this concept, showing how relative count rates on the scintillators vary dramatically with the direction of the GRB. GALI thus provides accurate directional information about the GRB source without compromising the wide field of view.
 In a sense, GALI is an all-sky 3D coded mask. 
A summary of the comparison between Swift-BAT, Fermi-GBM, and a (potential) GALI instrument is presented in \autoref{tab:comparison}. 
 GALI is still in the Research \& Development phase and is exploring a few launch opportunities. One of these options is the Czech QUVIK mission \cite{Werner_2024}. Some relevant simulations are presented in \autoref{sec:simulations}.

\begin{table} [ht]
\centering
\begin{tabular}{ |p{3cm}||p{3.4cm}||p{3.2cm}||p{4cm}|} 
 \hline
 Mission/ \break characterizations & Swift-BAT & Fermi-GBM & GALI in potential\\
 \hline
 Method     & coded mask aperture  & dispersed detectors & coded mask aperture from dispersed detectors \\
 Detectors    & 32,768 CdZnTe (CZT)   & 12 NaI(Tl) + 2 BGO & 362 CsI(Tl)\\  
 Energy range (keV)   & 15-150  & 8-40,000 & 10-700  \\
 Field of view (sr)      & 1.4 & 9.5 & 6.28\,* \\
 Angular accuracy (degrees)   & 0.016-0.06  & 3-15** & 2-5 **\\
 \hline
\end{tabular}
\caption{\label{tab:comparison}Summary comparison between Swift-BAT, Fermi-GBM and GALI.\\ *Depends on the TBD platform. ** For a range of detectable fluxes (\autoref{fig:GALIs Angle estimation error distribution}).} 
\end{table} 

\section{Building GALI}
\label{sec:gali}
\paragraph{} This section introduces the GALI directional detection method along with its structure in detail. Several key components and features that contribute to the performance and calibration process are discussed. 
%Some of these features were tested and adjusted to understand how they affect GALI performance. 

\paragraph{} The current GALI structure is composed of 362 (9mm)${^3}$ CsI(Tl) crystal scintillators arranged randomly on nine electronic boards. 
Each scintillator is coated with SiO${_2}$, wrapped with a reflective polymer foil, and coupled to a silicon photo-multiplier (SiPM) by a space-qualified encapsulant. 
Each detector unit - wrapped crystal with SiPM - is randomly coupled on a printed circuit board (PCB) with Arlon dielectric substrates.
 These random locations were selected following simulations that aim to achieve the best angular localizing accuracy (see Ref. \cite{Rahin2020}). 
The nine PCB boards are connected to six TOFPET ASICs (application-specific integrated circuits). Each SiPM is read by one ASIC channel. A multiprocessor system on chip with an FPGA and a computer is connected to read all the data from the ASICs. 

\subsection{Detection Method}
\paragraph{} 
%GALI has 362 cubic scintillators, each mounted on a silicon photo-multipliers (SiPMs) that measures the amount of light produced by ionizing radiation in the scintillator crystal \cite{Bourland2016}. 
The GALI method is to detect only a few source photons above the background in many scintillators, the ratios of which determine the GRB direction.
For this GALI needs to work at a low noise level and be sensitive to soft $\gamma$-rays typical of GRBs.
Although each scintillator receives only a few source photons, those can be significant when that individual scintillator is exposed to a small part of the sky and thus records negligible background.
 Triggering and spectra are obtained from the collective signal of all scintillators.

%\textcolor{red}{Julia, I added the above paragraph, instead of all the following text, which I commented out because it appears again under the components.}
%The resulting number of visible photons in the scintillator is proportional to the absorbed primary $\gamma$ photon energy \cite {Kramar2017}, which provides GALI also with the GRB spectra.
%The SiPMs are multi-pixel photon counters connected in parallel \cite{Eckert2010}, which amplify the signal with operation voltages of 40-43\,V; low compared to traditional photo-multiplier tubes (PMTs with 100s of Volts). The SiPMs are connected to the ASICs, which amplify and digitize the analog SiPM signals. 
%The PETsys ASICs are optimized for PET-positron emission tomography, a clinical time-of-flight imaging technique built for high data rates and fast timing of 511\,keV photons \cite{schug2019}, and not for GRB detection, which requires 
%%. Since GALI utilizes the relative counts detected between the scintillators to localize the source of the GRBs with high accuracy, it is optimized 
%sensitivity to low-energy $\gamma$ photons, for the most part at a low rate.
%Therefore, \textcolor{red}{we use the high voltage end of the SiPMS, and } several $\gamma$-ray lab sources to test the resulting signals from sources down to 10\,keV.

\subsection{GALI Components}
\vspace{0.2cm}
\begin{wrapfigure}{r}{0.45\textwidth}
    \includegraphics[scale=0.06]{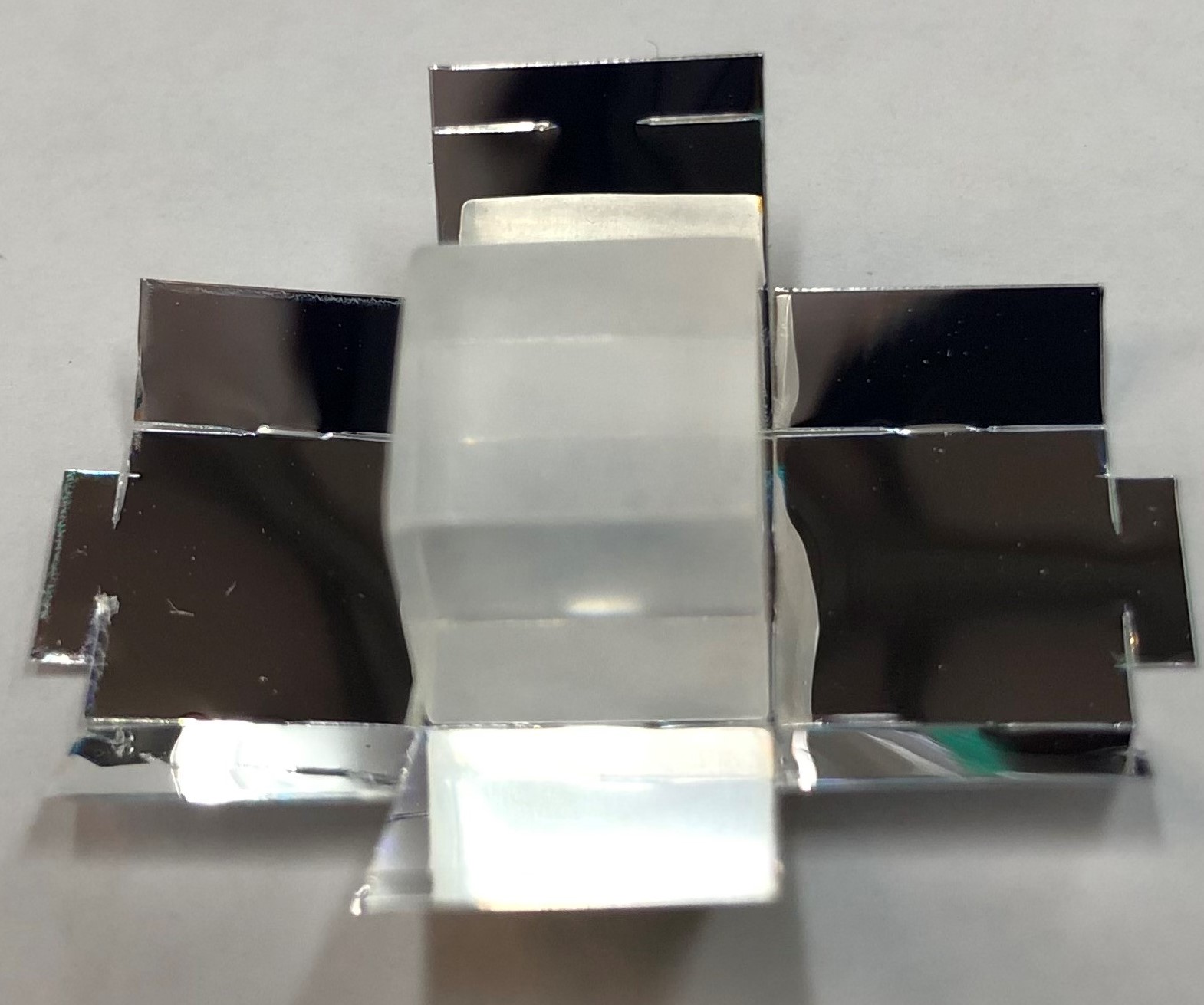}
  \caption{\label{fig:viquity} CsI(Tl) scintillator wrapped with Viquity.}
\end{wrapfigure}

\underline{\textbf{CsI(Tl) Cubic Scintillators}} \hspace{0.3cm} Thallium activated Cesium Iodide (CsI(Tl)) is chosen for its bright scintillation output of 54 photons/keV, with no self-activity and high detection efficiency.
 However, they have a relatively long decay time of $3.5$\,$\mu$s  \footnote{\url{https://www.berkeleynucleonics.com/cesium-iodide}.}, which hinders detection at high $\gamma$-ray rates due to pile-up, and is a poor match to the fast ASIC integration time $1.25\,\mu s$.
CsI is somewhat hygroscopic, hence each cubic (9\,mm)$^3$ crystal is coated with a thin layer of ${\sim}200\,\mu m$ of SiO${_2}$. In addition, each scintillator is wrapped in a reflective polymer - Viquity, see \autoref{fig:viquity}, to maximize internal reflection and keep stray light out.

\begin{wrapfigure}{r}{0.3\textwidth}
    \includegraphics[width=0.3\textwidth]{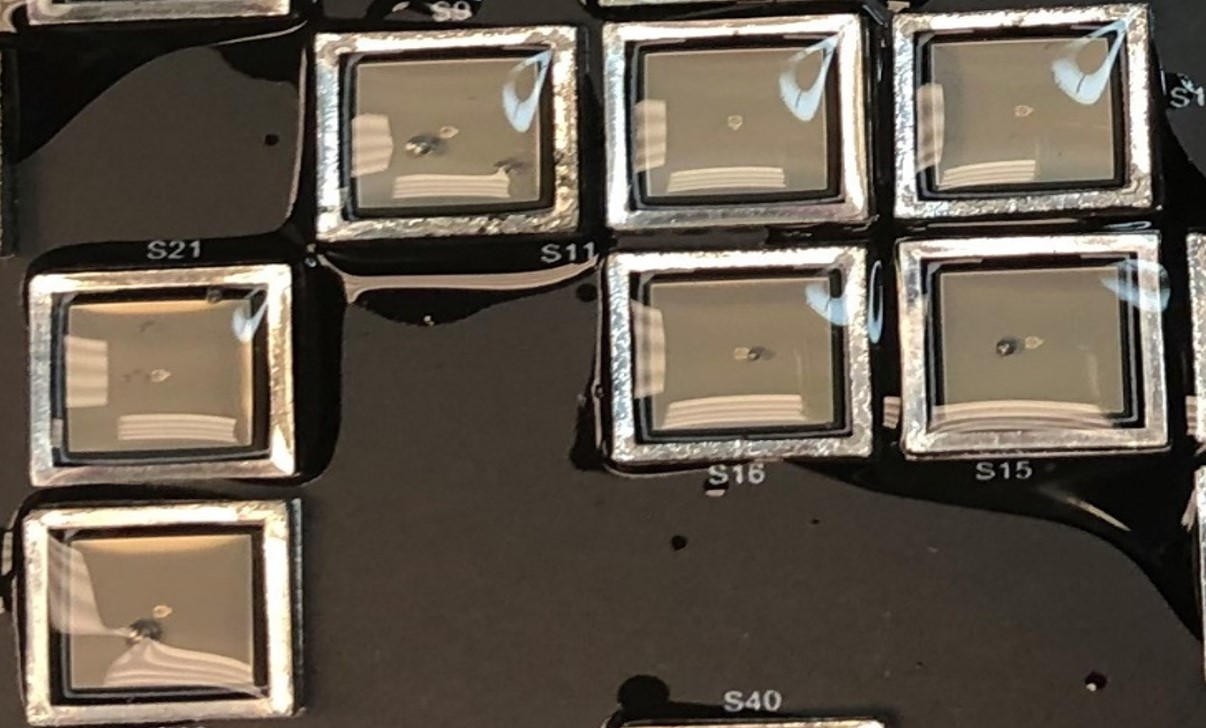}
  \caption{\label{fig:sipms}SiPMs soldered on Arlon boards with space grade encapsulation on the surfaces. The copper frame around each SiPM protects it from space radiation.}
\end{wrapfigure}

\newpage
\underline{\textbf{Silicon Photo-Multipliers (SiPMs)}} 
\hspace{0.3cm} The SiPMs receive the scintillation and produce an amplified current. For GALI, Hamamatsu S14160-6050HS SiPMs are used. 
%They have several features that distinguish them from other typical PMTs. E.g., operation in low voltage and high efficiency in photon detection. 
%The SiPM parameters were tested and adjusted to follow specific requirements \textcolor{red}{meaningless ???}.
 We use the SiPMs at over-voltage 4.5-5\, V, higher than the recommended value, 2.7\, V, to amplify the weak signal of low-energy $\gamma$-photons so it can be detected by the ASICs.

%\begin{table} [ht]
%\centering
%\begin{tabular}{ |p{6cm}||p{3cm}||p{3cm}|} 
% \hline
% Parameter & Symbol & Value With Units \\
% \hline
% Breakdown voltage  &$V_{BR}$  &38 V  \\
% Recommended operating voltage     &$V_{OP}$ &$V_{BR}+2.7 $ V \\
 %Gain      & M & $2.5 \times 10^6$\\
 %Peak sensitivity wavelength       &$\lambda_P$ &450 nm\\
 %Photon detection efficiency at $\lambda_P$        & PDE & 50\%    \\
 %Operation temperature &$T{opr}$ & -40$^{\circ}$C to +85 $^{\circ}$C\\
 %\hline
%\end{tabular}
%\caption{\label{tab:param}Parameters of  Hamamatsu S14160-6050HS SiPMs}
%\end{table}

The SiPMs are soldered to the electronic boards.
 The scintillators are then glued to the SiPMs with a DOWSIL$^{TM}$ 1200 OS primer and a 93-500 space-grade encapsulant (see \autoref{fig:sipms}).
 This process is done manually in a clean room and requires time and accuracy.
 First, surfaces are cleaned, then the primer is applied and given ample time to dry. 
While waiting, a two-component encapsulant must be mixed in a vacuum tube and then applied to the SiPMs. 
To complete the curing process, the whole layer goes into an oven at 30$^{\circ}$\,C for 24 hours. Thermal cycle tests will be conducted to ensure matching thermal expansion values with the scintillators and SiPMs. 

\begin{wrapfigure}{r}{0.3\textwidth}
    \includegraphics[width=0.3\textwidth]{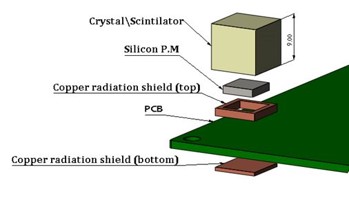}
  \caption{\label{fig:copper} Copper shields for each scintillator for radiation protection.}
\end{wrapfigure}

To protect the SiPMs from radiation damage in space, we also weld a copper frame on the upper side of the board, and where there is no scintillator below, a square copper panel on the other side of the boards (see \autoref{fig:copper}). 
Additionally, to minimize vibrations on the crystals during launch, 3M$^{TM}$ was double coated in the gap between them and the next board. 

\underline{\textbf{PCB Boards with Arlon dielectric substrates}} \hspace{0.3cm} GALI uses (nine) polyimide non-woven aramid 85NT- Arlon boards instead of standard (Si-based) PCBs to minimize $\gamma$-ray absorption in the boards.
 These are low in-plane $(X, Y)$ boards, i.e. reduced thickness, while retaining their robust properties under thermal variations.
 Each board is covered with a black mask to minimize stray light reflections onto the SiPMs, which results in unwanted electronic noise (see \autoref{fig:sipms}).
 Each board has 5 aluminum pillars, four at the corners and one in the middle to strengthen the mechanical structure, and to transport heat between the detector layers, and down to the satellite.

\vspace{0.2cm}
\underline{\textbf{Read-Out Electronics - ASICs}} \hspace{0.3cm} The 362 scintillators are read by six TOFPET2 ASICs. The ASICs are designed and evaluated for time-of-flight measurement in Positron Emission Tomography (TOF-PET) applications, not astrophysical ones. 

%In other words, it is optimized to measure the distance between a sensor and a LYSO crystal \textcolor{red}{no, between the point of emission and two opposite detectors 180$^\circ$ to each other}, based on the time difference between the emission of a signal from Na-22, and its return to the sensor after being reflected by the crystal.
%The default detecting process is very fast compared with the detection with CsI(Tl) crystals, and it is evaluated to detect 511\,keV line, and not particularly small energy lines as needed for localizing. 
%\textcolor{red}{Julia, this is too much, and I am not entirely sure there is any reflection "of a signal?" involved.}

%\textcolor{red}{Instead I suggest we write as follows:}

Although PET tomography is also aimed at localizing the $\gamma$-ray source, there are important differences concerning GRB localization. PET readout is designed to detect two 511\,keV $\gamma$-photons emitted in opposite directions, following positron annihilation. 
Timing precision is required to be better than ns, to localize the source, and to enable accurate tomography. 
In contrast, GRBs are dominated by soft $\gamma$ photons of 10's of keV, with no timing requirement, since the count rates are low, typically $<100,000$\,cts/s.
In GALI, these count rates are further distributed over dozens of detector units.

Consequently, using the PETsys ASIC for GRB detection is a challenge and it needs to be well calibrated for our requirements of detecting low-energy photons with the long decay times of the CsI(Tl) crystals. The ASIC evaluation kit consists of one FEM128 front-end module connected via flat cable to a front-end board FEB/D with an Ethernet link to the computer. The FEM128 has two ASIC boards reading 64 channels each. Activity in one channel does not result in any dead time in the other channels, allowing parallel reading. The kit is provided also with firmware and software \footnote{\url{https://www.petsyselectronics.com/web/}}.

\paragraph{} Each SiPM is read by a single ASIC channel, and since each ASIC reads 64 SiPMs, an adequate event reading rate is needed. The PETsys ASIC can read a million events per second, which is sufficient in almost all GRB detections. Each channel contains independent amplifiers, discriminators, time-to-digital converters, and charge-to-digital converters. The ASIC can work with two calibration modes: QDC mode - Charge to Digital Converter, and TDC mode - Time to Digital Converter. In this work, the ASIC operates in QDC mode which measures the integrated charge from the rising edge of the charge trigger until the end of the integration window.

\begin{figure}[h]
\centering
    \includegraphics[width=0.78\textwidth]{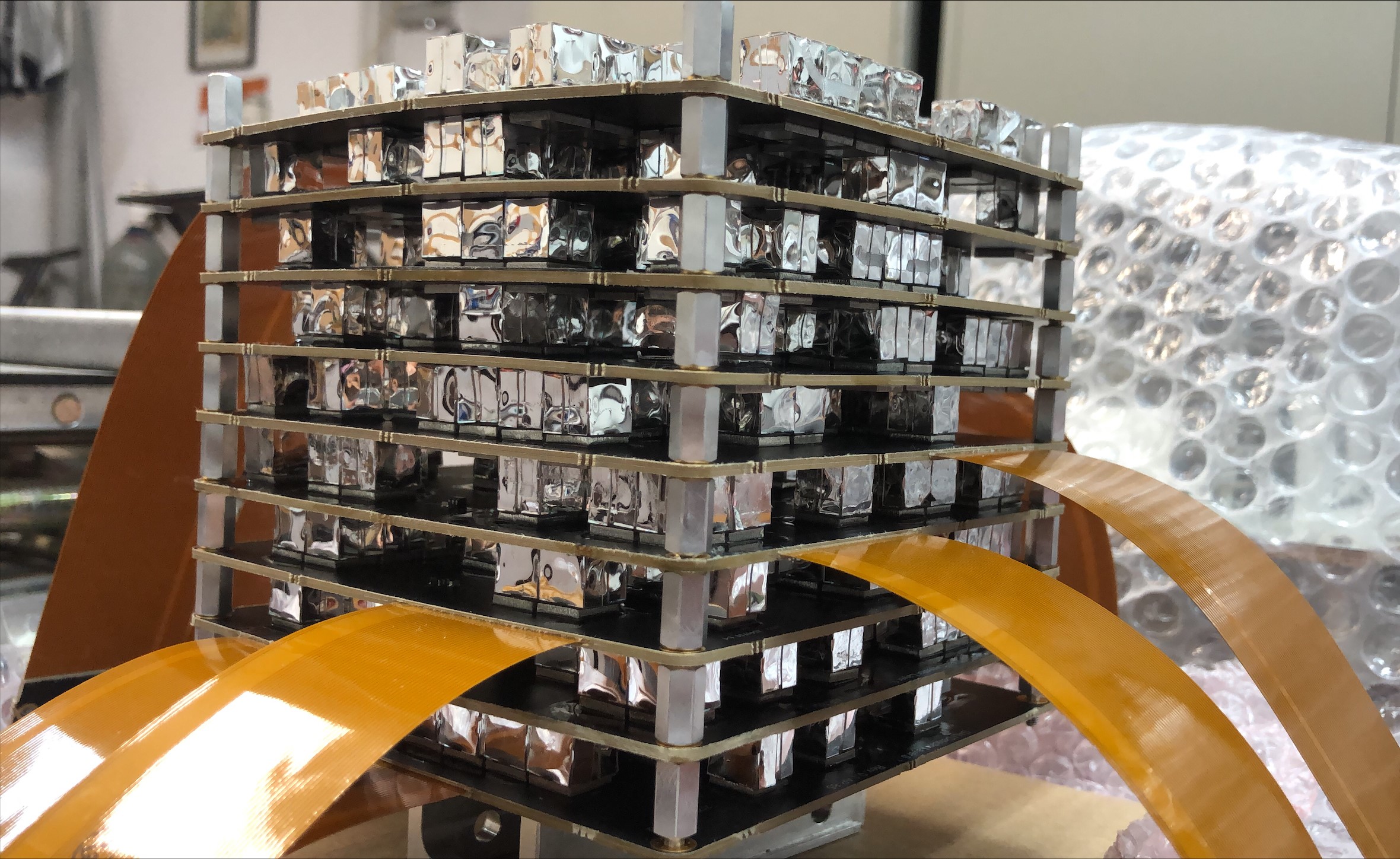}
\vspace{0.5cm}    
\caption{\label{fig:gali} GALI - The GAmma-ray burst Localizing Instrument, composed of 362 wrapped CsI(Tl) scintillators attached to 9 PCB printed electronic boards. Each board is printed for specific SiPM locations and is connected with flat cables to the ASICs and computer that sit underneath the detector unit, in an electronics box.}
\end{figure}

Finally, assembling all of these components, and then integrating and testing their electronics resulted in the new GALI structure, which is depicted in \autoref{fig:gali}. 

\section{Monte-Carlo Simulations}
\label{sec:simulations}

\begin{wrapfigure}{r}{0.3\textwidth}
    \includegraphics[width=0.3\textwidth]{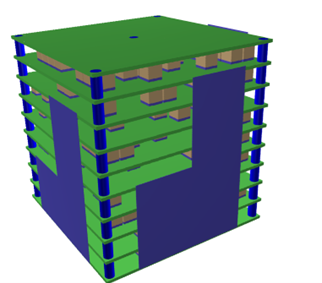}
  \caption{\label{fig:GALI_model}Simulation model of the current GALI detector - An 11x11x9 3D array.}
\end{wrapfigure}

\paragraph{} In order to further study the localization capabilities of GALI, a Monte-Carlo simulation of the detector was created. Background radiation is included as expected in low earth orbit (600\,km). %as we would expect to see on the QUVIK satellite \cite{Werner_2024}. 
The simulation was built using MEGAlib \cite{ZOGLAUER2006629}, a software library based on the Geant4 simulation toolkit \cite{AGOSTINELLI2003250, 1610988, ALLISON2016186}. 
The model of the detector unit includes all of the parts that can absorb or reflect $\gamma$-rays (See \autoref{fig:GALI_model}), including the scintillators and their thin metal cover, SiPMs and their copper shielding, the plastic layers and the aluminum pillars holding them together, the copper wiring covered in plastic casing on the sides of the detectors, and a thin metal cover in place of the thermal blanket the detector would be covered by in space.

\begin{figure}[h]
\centering
\includegraphics[width=0.8\linewidth]{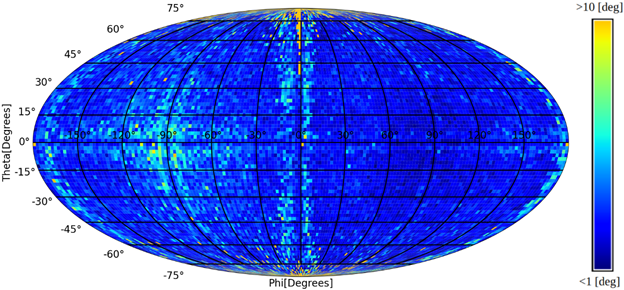}
\caption{\label{fig:A sky map of GALIs errors}Sky map of the angle estimation uncertainty of GALI over the full sky (no satellite platform assumed) based on our Monte Carlo simulations for 
%mapped over the estimated angle of the GRB, for 
GRBs with a photon flux of 10\,photons\,cm$^{-2}$\,s$^{-1}$ and duration of 1\,s with energy distribution of the Band function, between 10 and 1000 keV.}
\end{figure}

In this manner, the detector is simulated with the continuous background radiation, and with high-count GRBs (1000 photons\,cm$^{-2}$) on a sky grid to create a reference map of relative count rates on each scintillator, as a function of sky angle. 
Subsequently, the simulation generates numerous, much-fainter GRBs arriving from random directions, with a uniform photon flux of 10\,photons\,cm$^{-2}$\,s$^{-1}$ and a duration of 1\,s, with the Band broken power-law spectrum between 10 and 1000 keV.
The number of counts in each individual scintillator is registered for each GRB, and the angle of arrival is calculated by comparison with the said reference map using the algorithm described in \autoref{subsec:method}. 
The statistics of these numerous burst localizations is used to produce a map of GALI's average estimated angle uncertainty, which is presented in \autoref{fig:A sky map of GALIs errors} for $1.2\times10^5$ such GRBs. %mapped over the estimated angle, using the prior simulated calibration and background measurements. 
The consistently low uncertainty of a few degrees over the entire sky is evident.
%, and for the half-sky that we expect to be not obstructed (Right side of the map), we see that the average estimation error is consistently small over all of it. 
The left side of the sky in the figure is the satellite platform where GALI would be placed upon. Although no satellite is simulated here, the uncertainties there are larger due to the copper shielding underneath the SiPMs absorbing source photons (\autoref{fig:copper}).
%, as seen in the increase in error at [0,-90\,$^\circ$].

\begin{figure}[h]
\centering
\includegraphics[width=0.7\linewidth]{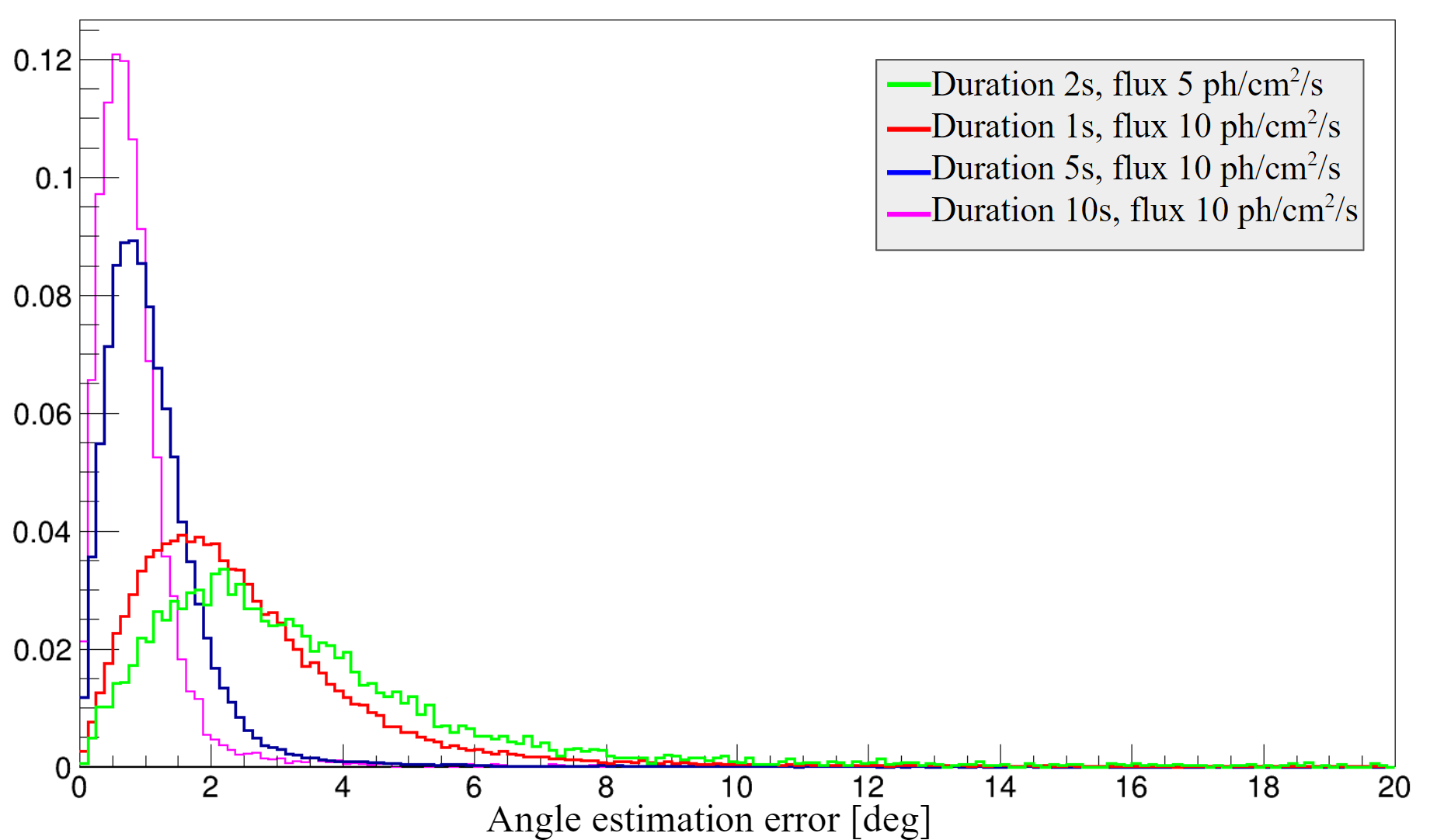}
\caption{\label{fig:GALIs Angle estimation error distribution}Angle estimation uncertainty distribution from GALI simulations in the energy band of 10-1000 keV, for different GRB fluxes and duration from the half sky above the detector.}
\end{figure}

In the same manner, GRBs were simulated with different fluxes and durations, from the half sky above the detector. The distribution of GALI angle estimation uncertainty for these GRBs is shown in \autoref{fig:GALIs Angle estimation error distribution}. 
For all of these cases, the uncertainty is contained within several degrees.
As expected, localization accuracy improves with the number of source photons, or fluence (flux times duration). For a fixed source fluence (e.g., green and red curves in the figure) it decreases with duration, as the source to background ratio decreases. 

\section{Experimental Setup and Localization Method }
\label{sec:localization}

\subsection{Experimental Setup}
\begin{figure} [h]
    \centering \includegraphics[scale=0.08]{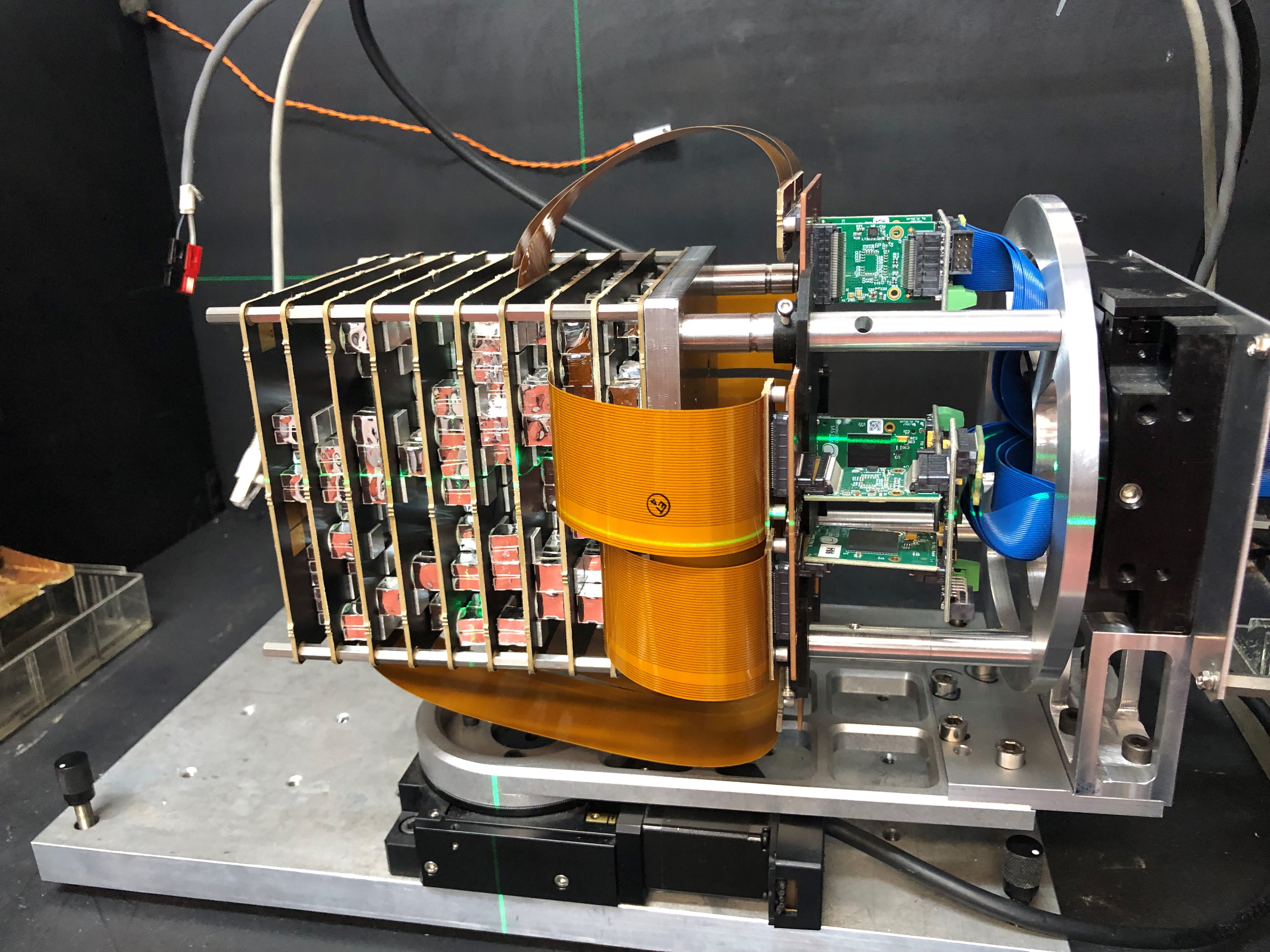} 
    \vspace{0.5cm}
\caption{\label{fig:gali_rot_sys} 2-axis rotation platform for GALI ocalization measurements. The source is located 4.5\,m to the left side of the image along the $\phi$-rotation axis.Each Arlon board is read out via one ASIC (on the right), which is placed opposite the source direction. All ASICs are connected via the blue cables to the FEB module and from it to the computer. The whole system is placed inside a thermally and optically isolating box.}
\end{figure}

\paragraph{} The assembled GALI detector was placed on a rotating platform that is connected to two rotation motors - 8MR190-2-28/30 (see \autoref{fig:gali_rot_sys}), which can rotate in 360\,$^\circ$.
One is defined to be $0\,^\circ\leq\theta\leq90\,^\circ$ (latitude) and the second to be $0\,^\circ\leq\phi\leq360\,^\circ$ (longitude) to cover a hemisphere. 
The motors are programmed to rotate in 5\,$^\circ$ intervals in both $\theta$ and $\phi$.
To avoid precession, we made sure the line going through all board centers coincides with the $\phi$-rotating axis. Furthermore, GALI weighs $\simeq$ 2\,kg, thus, it is expected to %have a small 
tilt. 
To calibrate these mis-alignments, we used a 2D laser scan to tune the detector $\phi$-rotation and the source axis, to be parallel to the platform. 

\paragraph{} The connection of the nine Arlon boards to the ASICs and FEB module underneath the detector unit is seen in  \autoref{fig:gali_rot_sys}. 
This setup uses five cables since each connector in the module reads two ASICs.
%The five cables go through a hole in the center of the rotating motor because the module sits behind it (see \autoref{fig:gali_rot_sys}). 
The FEB modules are connected to a computer, where the ASICs are controlled via a GUI interface.
The whole system - the detector, and the ASICs with its modules, are placed inside an optically and thermally isolated box, where temperature can be controlled. 
The power supply is placed outside the thermal box to reduce electromagnetic noise. 

%    \vspace{-1cm}

\subsection{Localization Experiment}
\paragraph{} GALI was exposed to a 10\,mCi $^{241}$Am source positioned approximately 4.5 meters away, to simulate a distant source and to test the direction reconstruction performance.
At this distance, the effective flux in the 59.6\,keV line is about 52\,ph\,cm$^{-2} $s$^{-1}$.% (as illustrated in \autoref{fig:simulatuins}). 
 Two different types of measurements were conducted at each 5$^\circ$ step covering the hemisphere. Long 120\,s exposures were used to calibrate the relative count rates expected on each scintillator and produce the reference map. Subsequently, 100 short 0.5\,s bursts were measured and their location was estimated from the said map. 
In each measurement and for each detector unit, we identified the spectral line at 59.6\,keV, and fitted it with a Gaussian to obtain the total measured counts. 
Overall, $(90/5)\times(360/5)+1=1297$ measurements were done to cover the hemisphere ($0\leq\theta\leq90, 0\leq\phi\leq 355$) in 5$^\circ$ steps.

\subsection{Localization Reconstruction Method}
\label{subsec:method}

\paragraph{} The GALI localization process is based on a statistical comparison between the measured and expected count rates for each scintillator. Since a low count rate (for the 100 short bursts) is measured on each scintillator, and the background is negligible in the lab, the measurements follow a Poisson distribution, $P(n;\lambda)$, where $n$ is the number of events, and $\lambda$ is the average. 
For a real GRB that varies in time, $\lambda$ is the average number of photons received from the source per unit time - count rate -, and $P(n;\lambda)$ will describe the probability of recording $n$ photons in a given exposure time $t$. %In the planned experiments we will only use the total count rate.  

\paragraph{}To study the maximum probability of the observed data given a model and its parameters, the maximum likelihood method is used, with $\theta$ and $\phi$ being the parameters that maximize the probability of the observed data. For a Poisson distribution, it is defined as the product of individual Poisson probabilities computed in each bin $i$,

\begin{equation}\label{eq:likelihood}
  L=\prod_{i}^{N} \frac{t m_i^{S_i}}{S_i!} e^{t m_i} \, ,
    \end{equation}
where $S_i$ are the observed counts, $t$ is the exposure time, and $m_i$ is the predicted count rate based on the model (i.e., reference map of 120\,s bursts). Since the experiment has no time dependence, this work only uses the total counts $tm_i$.

A Poisson log-likelihood function called Cash Statistic - Cstat \cite{Cash1979} is applied to calculate the localization accuracy in each angle interval. It can be derived from \autoref{eq:likelihood}, and defined as,  
\begin{equation}\label{eq:cstat}
  C=2\sum_{i=1}^{N} (tm_i)-S_i+S_i(\ln{(S_i)}-\ln{(tm_i)})\, .
    \end{equation}
%where it can be derived from \autoref{eq:likelihood} as follows, first, we take the natural logarithm of \autoref{eq:likelihood}, then use the Stirling approximation for $\ln{S_i!}\approx S_i\ln{S_i}-S_i $, and multiply by 2. Factor 2 ensures that changes in Cstat between models ($\Delta C$) will be distributed approximately as $\Delta \chi^2$ when the number of counts in each bin is high ($>$5)\footnote{\url{https://cxc.cfa.harvard.edu/sherpa/ahelp/cstat.html}}.

\paragraph{} In the present case, the model is defined by the calibrated reference map of the long bursts, and the short bursts are to be localized. 
To complete the calibration map we used a linear interpolation between the 5$^\circ$ intervals of the long bursts every 0.5\,$^\circ$. Subsequently, the Cstat value for each point on the calibration map was computed using \autoref{eq:cstat}. The minimal Cstat value represents the best estimate for the direction of the source. 
The angular distance %$\alpha$ 
of the 100 short bursts from the known position determines the uncertainty.
To obtain the average accuracy of GALI, we average this distance over all measured angles. 
%The accuracy of the interpolated grid points will be tested to find the optimal intervals. 

\subsection{Modeling the experiment with the Monte-Carlo simulation}
\paragraph{} In parallel to the experiment, a Monte-Carlo simulation of the experimental setup was carried out with MEGAlib. 
This simulation included the GALI detector as it was modeled in \autoref{sec:simulations}. The $^{241}$Am source is placed 4.5\,m away and with the same radioactivity as in the experiment.
Instead of the GALI being rotated however, the $\gamma$ point source was moved such that it was placed relative to the GALI detector in the same spots as it would in the experiment. Unlike previous simulations, this was done without any background radiation, as it models the experiment in the lab rather than in low earth orbit.

\section{ Localization Results}
\subsection{Experimental results}
\begin{figure}  [h]
    \centering
    \includegraphics[scale=0.4]{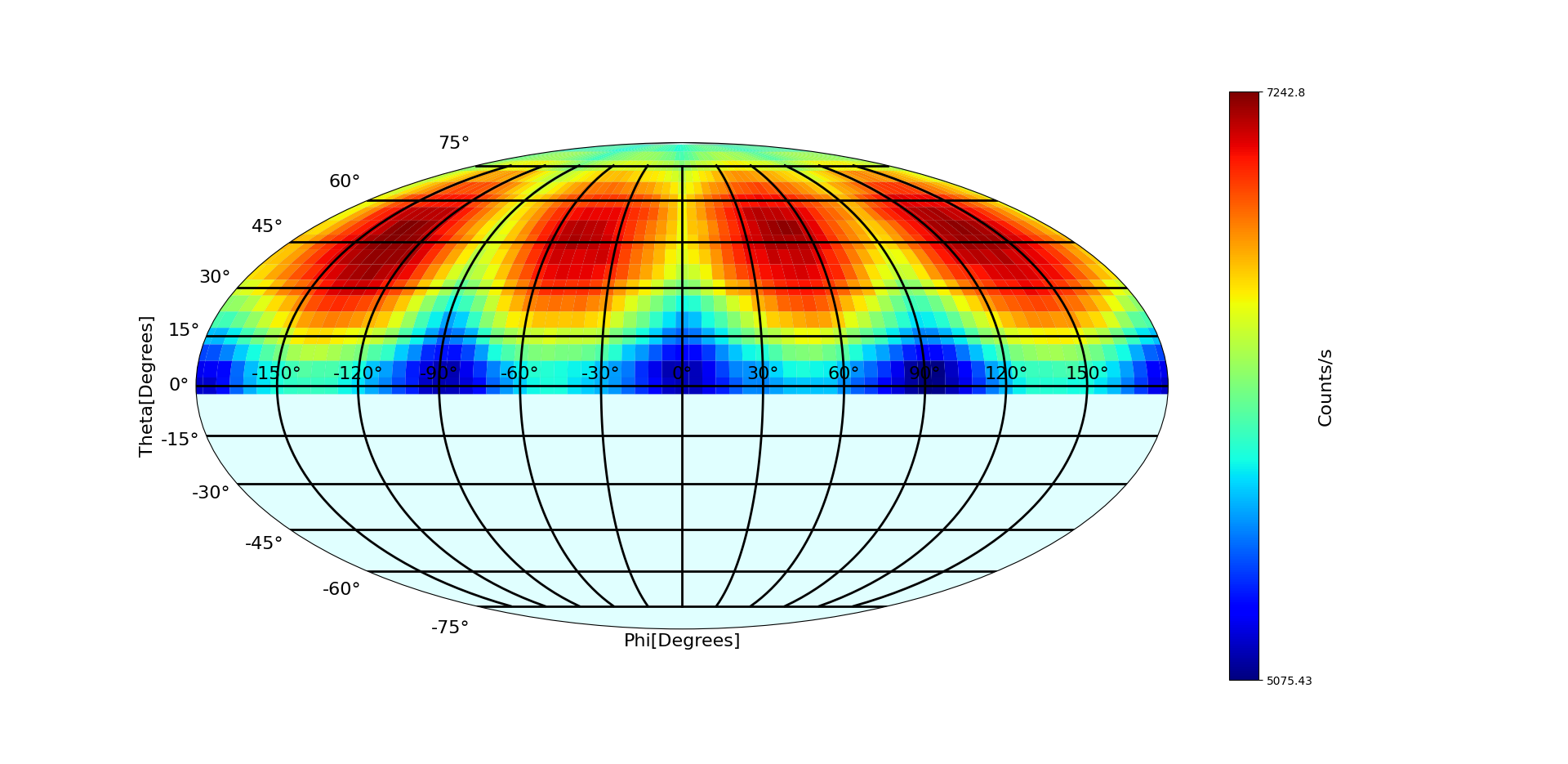} 
    \caption{Total count rates obtained from 120s exposures every 5 degrees. Colors represent the slight variation of the effective area of GALI with angle, peaking in the direction of the corners at 45$^\circ$. These measurements are used to calibrate the relative counts on each of the 362 scintillators, as a function of angle.}
    \label{fig:calibrationmap}
\end{figure}

%For each ($\theta,\phi$) in 5$^\circ$ intervals, 120\,s burst was measured, and the 59.6\,keV line was fitted to Gaussian curve for each scintillator to get the total counts. 
The results of our 120\,s exposures are presented in \autoref{fig:calibrationmap} in sky coordinates.
In this sky map, colors show the total count rates over the entire detector, and represent the slight variation of the effective area of GALI with angle. The red zones with the highest count rates ($\sim$ 7000\,cts/s) are from the corners of the detector where its geometrical cross section peaks, and the blue zones with the lowest count rates ($\sim$ 5000\,cts/s) are from face-on directions. Each box represents the 5$^\circ$ increments (1297 in total). 
To produce the calibrated reference map for localization measurements, we further linearly interpolated between the 5$^\circ$ intervals every 0.5\,$^\circ$.

\begin{figure} [h]
    \centering
    \includegraphics[scale=1.3]{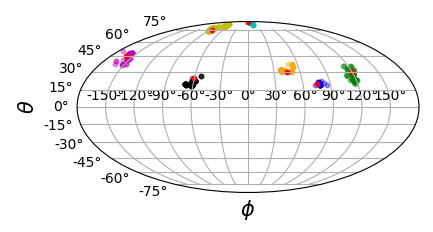}
    \caption{GALI directional reconstruction for seven arbitrary angular directions of ($\theta,\phi$) =\textcolor{blue}{(20,75)}, \textcolor{ForestGreen}{(30,120)}, \textcolor{GreenYellow}{(75,270)}, \textcolor{cyan}{(90,20)}, \textcolor{Mulberry}{(45,200)}, (25,300), and \textcolor{YellowOrange}{(30,45)} in sky coordinates measured in degrees. 
    %Each location is colored differently. 
    The red dot in each color cluster represents the true known position, while the colored dots represent the reconstruction of short (0.5\,s) bursts. The spread of the points around the original location shows the angular accuracy of GALI, which $<$2.3$^\circ$(90\% confidence) in all shown cases.}
    \label{fig:localization}
\end{figure}

 The localization measurements were carried out for 7 (arbitrarily chosen) ($\theta,\phi$) directions.
 At each direction we measured 100 0.5\,s short bursts, each producing 2500-3500 total photons on the detector. 
 Again, fitting the 59.6\,keV line to a Gaussian curve, we extracted the counts recorded in each scintillator.
 Subsequently, the Cstat value for each point on the calibration map was computed using \autoref{eq:cstat}.
 The minimal Cstat value represents the best localization estimate of the source. These results are presented in \autoref{fig:localization}.
%\autoref{fig:localization} shows the angular accuracy in localization with GALI. 
The red points represent the actual ($\theta,\phi$) position of the source. %, or the calibration position as measured from the 120\,s burst. 
The dots in each color cluster are the reconstructed positions of the 100 0.5\,s bursts from the same position. 
The scatter gives an idea of the uncertainty expected at each angular location.
For each cluster, we measured the mean values of the reconstructed ($\theta,\phi$), and the mean angular distance of the 100 dots from the original red position.
For the above measured sample, the angular accuracy radius was on average $<$2.3$^\circ$ (with 90\% confidence). 

\subsection{Simulation results}
\begin{figure}[h]
    \centering
    \includegraphics[scale=0.42]{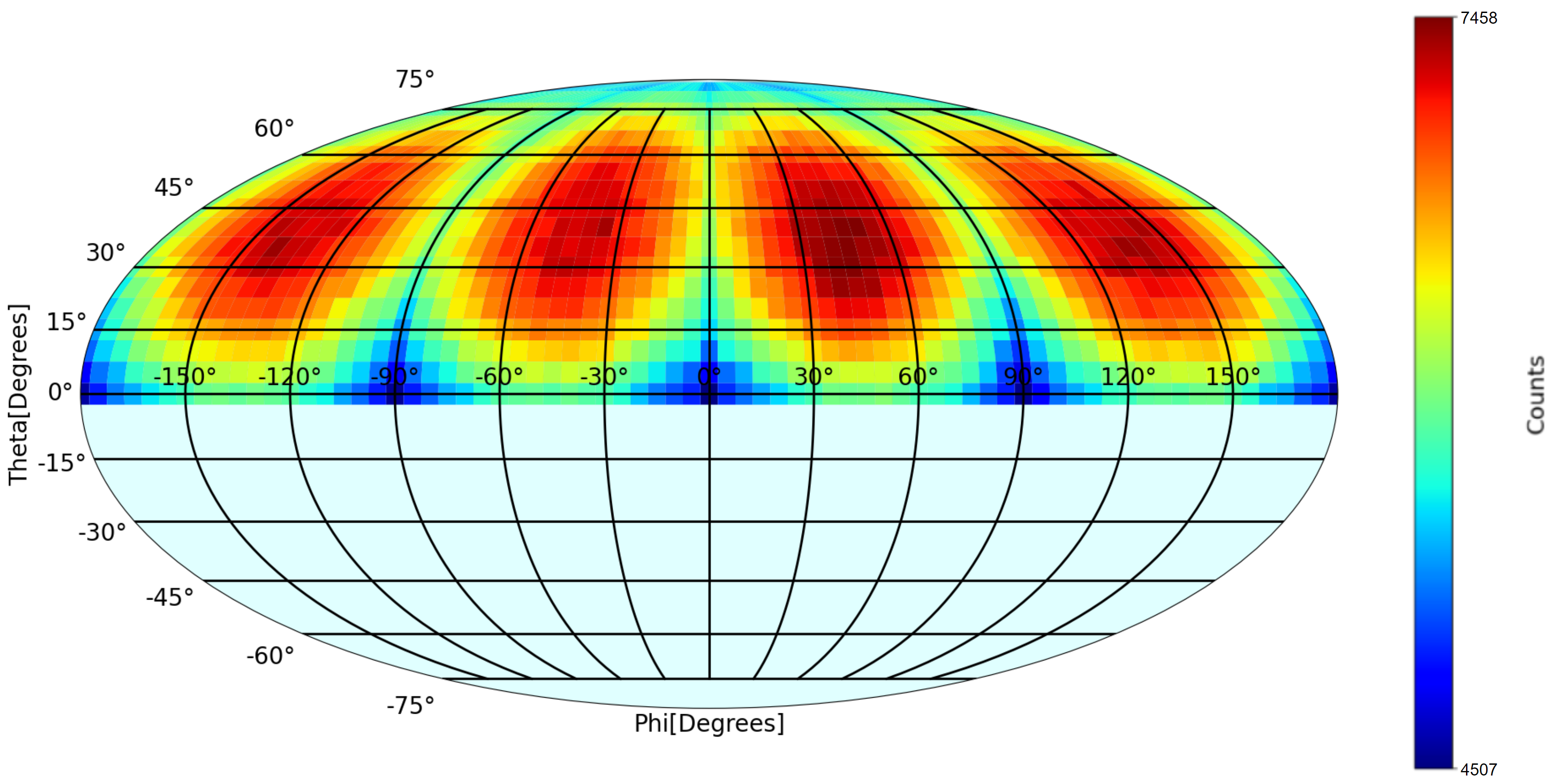} 
    \caption{
    Total counts registered in GALI in the Monte-Carlo simulation per second (In the 10-1000 keV energy band), for a source of 10\,mCi $^{241}$Am placed 450 cm from the center of the detector.}
    \label{fig:simcalibrationmap}
\end{figure}

\paragraph{} For each placement of the radiation source in the simulation, the number of total counts in the modeled scintillators was measured, and mapped in \autoref{fig:simcalibrationmap}. When comparing the two sky-maps, of the experimental (\autoref{fig:calibrationmap}) and simulated results, we see that the two are very similar in shape and count numbers. 
This stands as good validation for future simulations of GALI on a space platform with these tools, as it demonstrates its capabilities to accurately compute expected GALI count rates. %in the simulation space.

\section{Conclusion and Future Plans}
\label{sec:conclusions}
 
\begin{itemize}
    \item In this work, we present a novel $\gamma$-ray instrument, GALI, much smaller than existing GRB detectors, yet with potentially improved directional capabilities (\autoref{tab:comparison}).
    \item  A new structure (\autoref{fig:gali}), consisting of 362 scintillators obstructing $\gamma$-rays from each other, is demonstrated to have  exquisite angular radius reconstruction capabilities of $<$2.3$^\circ$ in the laboratory (with 90\% confidence) (\autoref{fig:localization}).
    \item This GALI configuration has an effective area of $\sim 100 - 140$\,cm$^2$, which is relatively uniform across the sky and peaks in the direction of the corners of the cube shaped detector (\autoref{fig:calibrationmap}, \autoref{fig:simcalibrationmap}).
%    \item  Between 30$^\circ$- 60$^\circ$, and particularly in 45$^\circ$, this GALI configuration has the highest count rate per angle, or the largest effective area. 
%    \item Angular reconstruction accuracy is being analyzed over the entire sky , to provide the expected accuracy of GALI over the full sky.
    \item The number of counts registered in the simulated scintillators was mapped over the sky, and the results are in excellent agreement with those measured in the lab. This gives us confidence that our simulation tools can reliably predict GALI's performance in space for real GRBs and sky background.
\end{itemize}

\acknowledgments % equivalent to \section*{ACKNOWLEDGMENTS}       
This work is partly funded by a grant from the Pazy Foundation.  J.S.N. acknowledges a doctoral fellowship from the Parasol Foundation Women in SpaceIL initiative.

%This unnumbered section is used to identify those who have aided the authors in understanding or accomplishing the work presented and to acknowledge sources of funding.  

% References
\bibliography{report} % bibliography data in report.bib
\bibliographystyle{spiebib} % makes bibtex use spiebib.bst

\end{document}